\documentclass[conference]{IEEEtran}
\IEEEoverridecommandlockouts
\usepackage{cite}
\usepackage{amsmath,amssymb,amsfonts}
\usepackage{algorithmic}
\usepackage{graphicx}
\usepackage{textcomp}
\usepackage{xcolor}
\usepackage{booktabs}
\usepackage{siunitx}
\usepackage{subcaption}

\usepackage{amsthm}
\usepackage{hyperref}
\usepackage[section]{placeins}
\usepackage{float}
\usepackage{tikz}
\usetikzlibrary{shapes,arrows,positioning,calc,fit}

\newtheorem{theorem}{Theorem} [section]

\theoremstyle{definition}
 
\newtheorem{remark}[theorem]{Remark}

\begingroup\expandafter\expandafter\expandafter\endgroup
\expandafter\ifx\csname pdfsuppresswarningpagegroup\endcsname\relax
\else
\fi

\def\BibTeX{{\rm B\kern-.05em{\sc i\kern-.025em b}\kern-.08em
    T\kern-.1667em\lower.7ex\hbox{E}\kern-.125emX}}

\begin{document}

\title{Quality of service based radar resource management using deep reinforcement learning
}

\author{\IEEEauthorblockN{Sebastian Durst}
	\IEEEauthorblockA{\textit{Fraunhofer-Institut f\"ur Hochfrequenz-} \\
		\textit{physik und Radartechnik FHR}\\
		Wachtberg, Germany \\
		sebastian.durst@fhr.fraunhofer.de}
	\and
	\IEEEauthorblockN{Stefan Br\"uggenwirth}
	\IEEEauthorblockA{\textit{Fraunhofer-Institut f\"ur Hochfrequenz-} \\
		\textit{physik und Radartechnik FHR}\\
		Wachtberg, Germany \\
		stefan.brueggenwirth@fhr.fraunhofer.de}
}

\maketitle

\begin{abstract}
	An intelligent radar resource management is an essential milestone in the development of a cognitive radar system. The quality of service based resource allocation model (Q-RAM) is a framework allowing for intelligent decision making but classical solutions seem insufficient for real-time application in a modern radar system.
	In this paper, we present a solution for the Q-RAM radar resource management problem using deep reinforcement learning considerably improving on runtime performance.
\end{abstract}



\tikzset{
	block/.style = {draw, fill=white, rectangle, minimum height=3em, minimum width=3em},
	tmp/.style  = {coordinate}, 
	sum/.style= {draw, fill=white, circle, node distance=1cm},
	input/.style = {coordinate},
	output/.style= {coordinate},
	pinstyle/.style = {pin edge={to-,thin,black}
	}
}


\section{Introduction}
\label{sec:intro}

The evolution of radar and antenna technology enables modern radar systems to simultaneously perform potentially conflicting functions, which is why resource management (i.e.\ the parameter selection, prioritisation and scheduling of tasks) is an essential part of any radar system and furthermore, an important building block of a fully cognitive setup (cf.~\cite{Haykin2006}).

\begin{figure}[!htb]
	\centering
	\begin{tikzpicture}[auto, node distance=2cm,>=latex', scale=0.7, every node/.style={scale=0.7}]
		\node [block, text width=3.5cm] (block0) {\small further processing: \textit{tracker update, task generation,\ldots}};
		\node [block, below of=block0, node distance=2.5cm, text width=1.4cm] (block1) {\small resource allocation};
		\node [block, right of=block1, node distance=4cm, text width=1.5cm] (block2) {\small scheduling};
		\node [block, right of=block2, node distance=4.0cm, text width=1.8cm] (block3) {\small job execution (antenna)};
		\node [block, above of=block3, node distance=2.5cm] (block4) {\small signal processing};
		\draw [->] (block0) -- node[text width=1cm]{\small tasks}(block1);
		\draw [->] (block1) -- node[text width=2cm]{\small selected jobs}(block2);
		\draw [->] (block2) -- node[text width=1cm]{\small timeline} (block3);
		\draw [->] (block3) -- node[text width=1cm]{\small received signal} (block4);
		\draw [->] (block4) -- node[text width=1cm]{\small processed signal} (block0);
		\tikzset{dotted/.style={draw=black, line width=1pt,
				dash pattern=on 3pt off 3pt,
				inner sep=4mm, rectangle}};	
		\node (dotted box) [dotted, fit = (block1) (block2), text width=5.3cm] {};
		\node at (dotted box.south) [below, inner sep=2mm] {\textbf{resource management}};
	\end{tikzpicture}
	\caption{A simplified radar architecture.} \label{fig:radar_architecture}
\end{figure}
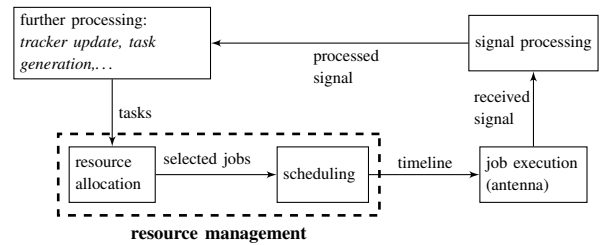

The system has a list of tasks to fulfil, each of which can be performed using a multitude of possible configurations differing in resource requirements and resulting utility.
The resource management module has to optimise the overall system utility while respecting global resource bounds and minimum quality of service requirements.
To this end, it chooses specific configurations for each task that are then aligned on the radar timeline by the scheduler
(see Figure~\ref{fig:radar_architecture}).
In particular, the complete problem is more involved than the sole scheduling of pre-defined tasks.
The problem was formulated under the name \emph{Quality of service based resource allocation model (Q-RAM)} and given a mathematical framework in \cite{Raj1997} and first applied to radar systems in \cite{Ghosh2004, Ghosh2006}. We will recall it in Section~\ref{sec:qram_problem}.
A (in general approximative) solution to the Q-RAM problem is given in \cite{Raj1997, Lee1998, Lee1999, Ghosh2004, Ghosh2006} and will be briefly outlined in Section~\ref{sec:classical_solution} as it motivates our own strategy.
Other notable approaches include \cite{CharlishPhD}. For a general introduction to radar resource management we refer the reader to \cite{MooDing}.

The formulation of the radar resource management problem via Q-RAM is without doubt as natural as it is powerful.
However, the authors are not aware of any implementation in productive radar systems as classical solution strategies do not seem to perform well enough in real-time.
The approach presented in this paper is based on deep reinforcement learning.
By designing a suitable environment, a neural network agent learns to predict a sequence of desirable task configurations without the necessity to embed all configurations into resource-utility-space.
This results in a drastically reduced computation time with only minor utility losses.
Applications of deep reinforcement learning or deep learning in general to resource management exist (cf.~\cite{Mao2016} or \cite{Shaghaghi2018} for radar). However, it is not the Q-RAM problem described in Section~\ref{sec:qram_problem} with the aim of optimising task configurations but the scheduling of fixed tasks that is investigated in the literature.
A notable exception is \cite{Dong2020}, where deep learning is used to optimise 5G radio networks under quality-of-service constraints but the chosen solution involving cascading neural networks and training with labelled data is very different from our strategy.
Further applications of reinforcement learning in radar focus on specific problems such as interference management, jamming and beamforming (cf.~\cite{Ak2019, Kang2018, Liu2019, Ahmed2020}).

\section{The Q-RAM problem}
\label{sec:qram_problem}

The Q-RAM problem is briefly introduced in this section.
Consider a multifunction radar.
Its goal is to maximise the \emph{utility} of a set of radar tasks over \emph{operational parameters} (e.g.\ waveform, dwell period, choice of tracking filter)
under \emph{resource constraints} (e.g.\ radar bandwidth, power) taking into account the \emph{environmental conditions}, i.e.\ situational data.
Here, utility depends on \emph{quality} which itself relates to the specific task configurations and environmental conditions
and allows for easier, interpretable user control.

Let $\{\tau_1,\ldots,\tau_n\}$ be a set of radar tasks and let there be $k$ types of resources with resource bounds $R_1,\ldots,R_k$.
Associated with each task $\tau_i$ are
\begin{itemize}
\item a discrete operational space $\Phi_i$, i.e.\ a discrete space of feasible task configurations,
\item a function $g_i\colon\thinspace \Phi_i\rightarrow\mathbb{R}^k$ mapping task configurations to their resource requirements,
\item a quality space $Q_i$ and
an environment space $E_i$,
\item a map  $f_i\colon\thinspace \Phi_i\times E_i\rightarrow Q_i$ associating a quality level to a configuration-environment-pair and
\item a quality-based utility function $\widetilde{u}_i\colon\thinspace Q_i\times E_i\rightarrow\mathbb{R}$.	
\end{itemize}
We define $u_i\colon\thinspace \Phi_i\times E_i\rightarrow\mathbb{R}$ via
$u_i(\phi, e) := \widetilde{u}_i(f_i(\phi, e), e)$
and the \emph{system utility} $u$ for chosen configurations
$\phi = (\phi_1,\ldots,\phi_n) \in \Phi := \Phi_1\times\cdots\Phi_n$
under environmental conditions
$e=(e_1,\ldots,e_n) \in E := E_1\times\cdots E_n$ as
$
u(\phi, e) = \sum_{i=1}^{n} u_i(\phi_i, e_i).
$
Now for fixed environmental data $e\in E$, the aim is to optimise global system utility while respecting resource bounds, i.e.\ we have 
the following optimisation problem:
\begin{align}
\max_{\phi = (\phi_1,\ldots,\phi_n)}& u(\phi, e)\\
\textrm{s.t. } \forall j=1,\ldots,k\;\;& \sum_{i=1}^n \big(g_i(\phi_i)\big)_j \leq R_j.
\end{align}

\section{The classical approach}
\label{sec:classical_solution}

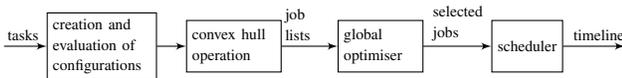
\begin{figure}[!htb]
	\centering
	\begin{tikzpicture}[auto, node distance=2cm,>=latex', scale=0.65, every node/.style={scale=0.65}]
	\node [input, name=input] (input) {};
	\node [block, right of=input, node distance=2cm, text width=2cm] (block1) {\small creation and evaluation of configurations};
	\node [block, right of=block1, node distance=2.7cm, text width=1.7cm] (block2) {\small convex hull operation};
	\node [block, right of=block2, node distance=3.0cm, text width=1.5cm] (block3) {\small global optimiser};
	\node [block, right of=block3, node distance=3.0cm] (block4) {\small scheduler};
	\node [output, right of=block4] (output) {};
	\draw [->] (input) -- node{\small tasks\;} (block1);
	\draw [->] (block1) -- (block2);
	\draw [->] (block2) -- node[text width=1cm]{\small job lists} (block3);
	\draw [->] (block3) -- node[text width=1cm]{\small selected jobs} (block4);
	\draw [->] (block4) -- node{\small \; timeline}(output);
	\end{tikzpicture}
	\caption{Block diagram of classic solution method.} \label{fig:classic_q-ram}
\end{figure}

A solution to the Q-RAM problem is proposed in \cite{Raj1997, Lee1998, Lee1999, Ghosh2004, Ghosh2006} and will be briefly outlined in the following.
If there are multiple types of resources $R_1,\ldots,R_k$, a so-called \emph{compound resource} is used, i.e.\ a function
$h\colon\thinspace\mathbb{R}^k\rightarrow\mathbb{R}$ mapping a resource vector to a scalar measure of resource requirements.
On a per task basis, all possible task configurations are generated and evaluated. This yields an embedding from the space of task configurations into resource-utility-space. A convex hull operation is used to determine the subset of configurations maximising utility for fixed resource levels.
We will refer to this subset of desirable configurations as a \emph{job list}.
A global optimiser then iteratively allocates resources to the task offering the best utility-to-resource-ratio provided sufficient resources are available.
After the resource allocation step, the resulting jobs are scheduled by a scheduler.
A block diagram of the process is given in Figure~\ref{fig:classic_q-ram}.

Although the ($2$-dimensional) convex hull operation can be performed in at most $\mathcal{O}(n\log{}n)$ with $n$ the number of points (i.e.\ possible task configurations), the embedding of configurations into resource-utility-space at runtime is costly.
Two possible solutions to improve runtime performance are proposed in \cite{Ghosh2006}: \emph{fast-traversal techniques} and \emph{quantisation of environmental data}. The first strategy reduces the number of configurations being created and evaluated,
the latter groups task into classes and uses pre-computed data.
Both methods rely on the inherent structure of the utility and resource functions used. As such, they can require thorough manual tweaking and domain knowledge.

Also note that
the classical Q-RAM solution is only of an heuristic nature, even when considering only a single resource, i.e.\ the calculated solution does not have to be optimal in general.
This is because a configuration point of a task in an optimal combination of task configurations is not necessarily lying on the boundary of the convex hull of all configuration points of that particular task in resource-utility-space because of resource restrictions
(also cf.\ Remark~\ref{rem:dep_config_space}).
Hence, a greedy algorithm only considering boundary points cannot produce the optimal solution in general.

\section{Approach using reinforcement learning}
\label{sec:our_solution}

\begin{figure}[htb] 
	\centering
	\includegraphics[width=.25\textwidth]{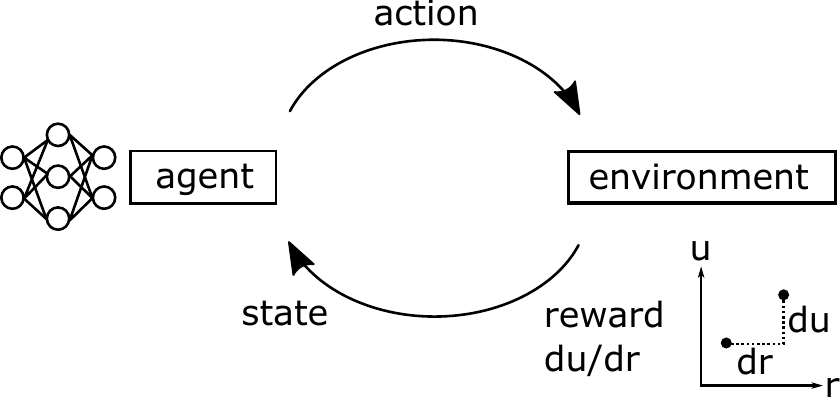}
	\caption{The agent interacting with its environment.}
	\label{fig:agent_env}
\end{figure}

Our approach to solving the radar resource management problem is based on deep reinforcement learning,
a branch of machine learning in which a software agent learns to take actions to maximise a cumulative reward signal by interacting with its environment (cf.\ Figure~\ref{fig:agent_env}).
In this section we will outline the general strategy without detailing particularities like specific network architectures, as these can be chosen in multiple ways depending e.g.\ on the size of the problem at hand.

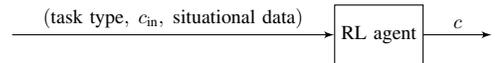
\begin{figure}[!htb]
	\centering
		\begin{tikzpicture}[auto, node distance=2cm,>=latex', scale=0.75, every node/.style={scale=0.75}]
			\node [input, name=input] (input) {};
			\node [block, right of=input, node distance=6.5cm] (block1) {RL agent};
			\node [output, right of=block1] (output) {};
			\draw [->] (input) -- node{$(\textrm{task type},\; c_{\textrm{in}},\; \textrm{situational data})$} (block1);
			\draw [->] (block1) -- node{$c$} (output);
		\end{tikzpicture}
	\caption{The agent.} \label{fig:agent}
\end{figure}

In reinforcement learning, the problem is modelled as a Markov Decision Process although not necessarily explicitly so.
The agent gets as input a state and outputs an action to perform in that state, i.e.\ it describes a policy for the Markov Decision Process (cf.~\cite{SuttonBarto}).
In our model a \emph{state} is a tuple consisting of some task configuration, task type and situational data, the latter can for example encode the tracker state for a tracking task.
The action space is made up of all possible configurations for the task under consideration.
We want the agent to output the task configuration that yields the highest difference quotient of utility to resource with the input configuration $c_{\textrm{in}}$, i.e.\ the configuration $c$ maximising
$
(u(c) - u(c_{\textrm{in}}))/(c - r(c_{\textrm{in}})),
$
where $u$ is a function modelling task utility and $r$ a scalar measure for resource consumption
(see Figure~\ref{fig:agent}).

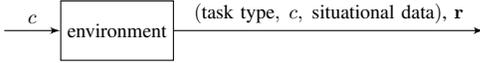
\begin{figure}[!htb]
	\centering
		\begin{tikzpicture}[auto, node distance=2cm,>=latex', scale=0.75, every node/.style={scale=0.75}]
		\node [input, name=input] (input) {};
		\node [block, right of=input] (block1) {environment};
		\node [output, right of=block1, node distance=6.5cm] (output) {};
		\draw [->] (input) -- node{$c$} (block1);
		\draw [->] (block1) -- node{$(\textrm{task type},\; c,\; \textrm{situational data})$, $\mathbf{r}$} (output);
		\end{tikzpicture}
	\caption{The environment.} \label{fig:env}
\end{figure}

The agent learns by interacting with its environment, which we have to model accordingly.
In particular, the environment incorporates the performance models and the functions to determine resource requirements and utility evaluation of a task configuration.
The environment receives the action chosen by the agent as input and outputs a new state, which is just the old state with the configuration component replaced by the chosen action, as well as a reward signal (see Figure~\ref{fig:env}).
This \emph{reward} $\mathbf{r}$ is based on the above utility-resource-quotient but can be adapted to increase training stability and performance (cf.\ Figure~\ref{fig:agent_env}).

Well-known reinforcement learning algorithms can be used to train the agent with randomly generated situational data using the environment mechanics described above.
Depending on the particular chosen network architecture this process can vary slightly, see e.g.\ \cite{Mnih2015, Lillicrap2015, Dulac2015}.
To summarise, the agent learns to output a beneficial next task configuration given some task configuration and situational data.

\subsection{Implementation of the agent in a resource management module}
\label{sec:method2}
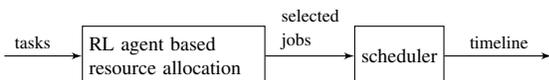
\begin{figure}[!htb]
	\centering
	\begin{tikzpicture}[auto, node distance=2cm,>=latex', scale=0.75, every node/.style={scale=0.75}]
	\node [input, name=input] (input) {};
	\node [block, right of=input, node distance=3cm, text width=3cm] (block1) {RL agent based resource allocation};
	\node [block, right of=block1, node distance=4cm] (block2) {scheduler};
	\node [output, right of=block2, node distance=2.7cm] (output) {};
	\draw [->] (input) -- node[text width=1cm]{\small tasks} (block1);
	\draw [->] (block1) -- node[text width=1cm]{\small selected jobs} (block2);
	\draw [->] (block2) -- node[text width=1cm]{\small timeline} (output);
	\end{tikzpicture}
	\caption{Block diagram of the implementation.} \label{fig:alternative_2}
\end{figure}

In the classical approach the global optimiser allocates resources iteratively by upgrading a task from a given configuration to the next configuration on the job list.
Hence, complete job lists are not actually required in the global optimisation but their items can be computed only when needed.
Observe that the described RL agent has exactly the capability to calculate the next desirable configuration from any given configuration.
The optimisation of resource allocation can thus be performed in a single module:
\begin{enumerate}
	\item The set of all tasks $T = \{t_1,\ldots,t_n\}$ is received as input and all tasks are set to a base configuration.
	\item The agent is applied to the base configuration $t_i^0$ of every task $t_i$ to obtain the next desired configuration $t_i^1$.
	\item The  utility-resource-quotient of the configurations
	is calculated for all tasks $t_i$.
	Here, the scalar compound resource is used in case there are multiple resources.
	\item The task $t_j$ with the highest ratio is set to configuration $t_j^1$ if sufficient resources are available. The agent is applied to this new configuration and the new utility-resource-ratio is determined.
	If an upgrade is not possible, the task with the next best ratio is tried to upgrade.
	\item The above step is repeated until upgrades are no longer possible, i.e.\ when resource bounds are reached.
\end{enumerate}
The resulting task configurations are then given to the scheduler to be placed on the radar timeline.
Note that the utility-resource-ratios can be weighted by the global optimiser to take into account task priorities.
A block diagram of the process is given in Figure~\ref{fig:alternative_2}.

\subsection{Implementation in radar systems}
\label{sec:implementation_radar}

It is worth noting that existing Q-RAM implementations can be retrofitted to the RL agent by simply replacing the convex hull operation by the trained network (cf.\ Section~\ref{sec:method2}).
Furthermore, such an integration allows
for using the classic Q-RAM convex hull computation in operation for task types not considered during training.
It is a unique feature of the presented reinforcement learning approach that the
system
can in principle be retrained with environment data collected
in the operational environment, e.g.\ by replacing the theoretically estimated track error by the real track error encountered during the mission.
This enables the ``self-learning" capability required for a cognitive radar system.

Moreover, the RL based agent resource management module can be easily ported to embedded and flight-certified hardware,
since only a standard feed-forward neural network has to be integrated on the airborne radar platform
and audited, the more complex training algorithm happens offline before the mission and does not have to be ported.

\section{Experimental results}
\label{sec:experimental_results}

\subsection{Scenario}
\label{sec:scenario}

\begin{figure}[htb] 
	\centering
	\includegraphics[width=.35\textwidth]{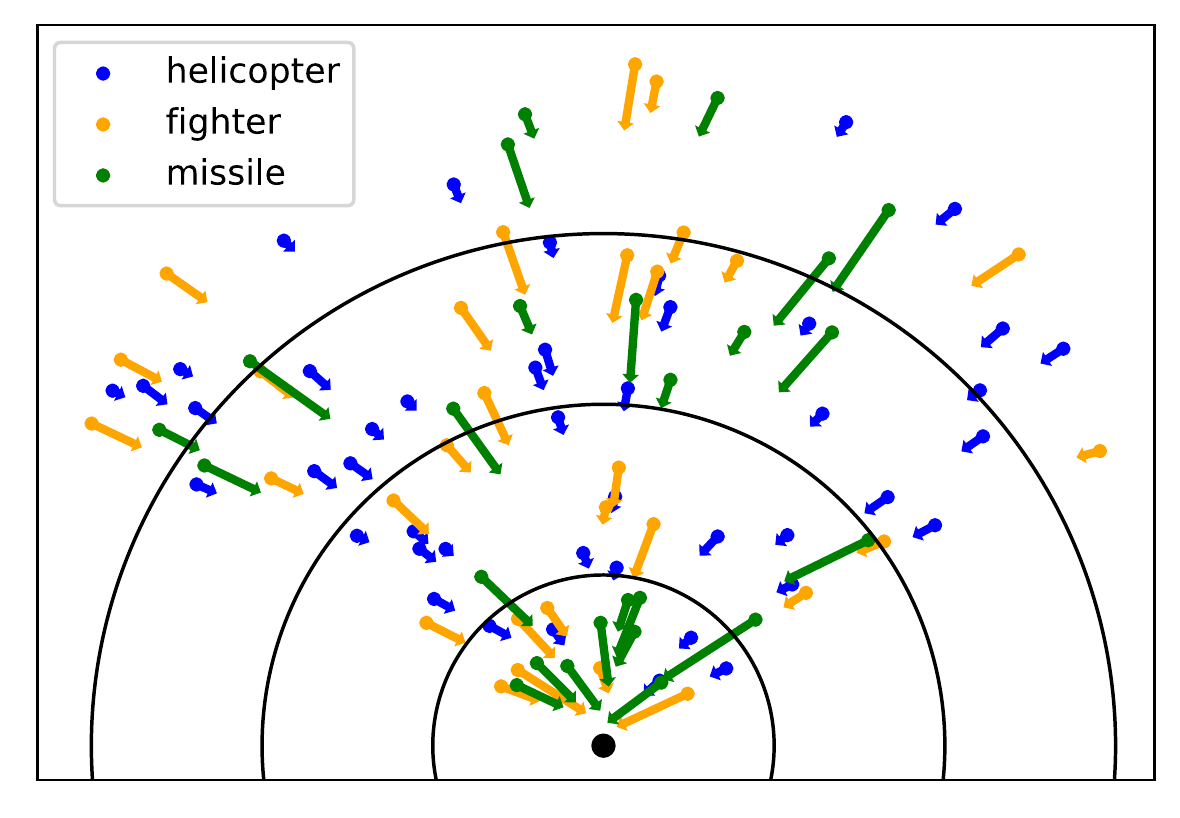}
	\caption{Speed and position of targets relative to the radar by target type in a randomly generated scenario.}
	\label{fig:scenario_plot}
\end{figure}

\begin{figure}[htb] 
	\centering
	\begin{subfigure}{.225\textwidth}
		\includegraphics[width=\textwidth]{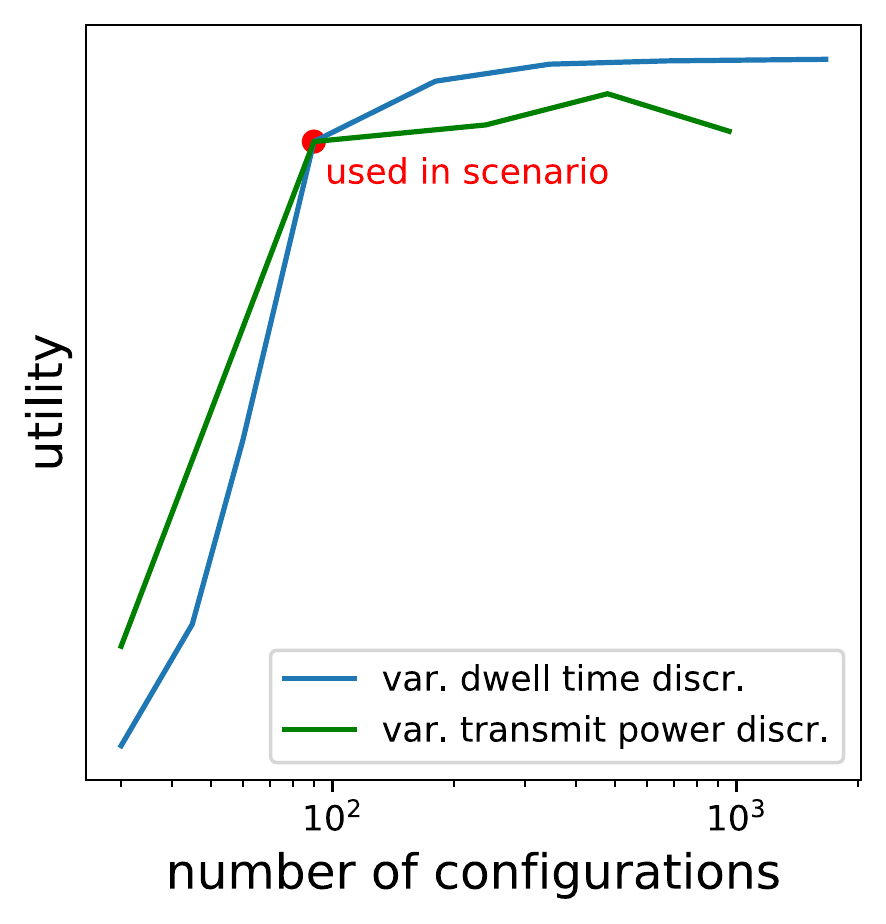}
		\caption{main scenario}
		\label{fig:util_by_configs_A_T}
	\end{subfigure}
	\begin{subfigure}{.225\textwidth}
		\includegraphics[width=\textwidth]{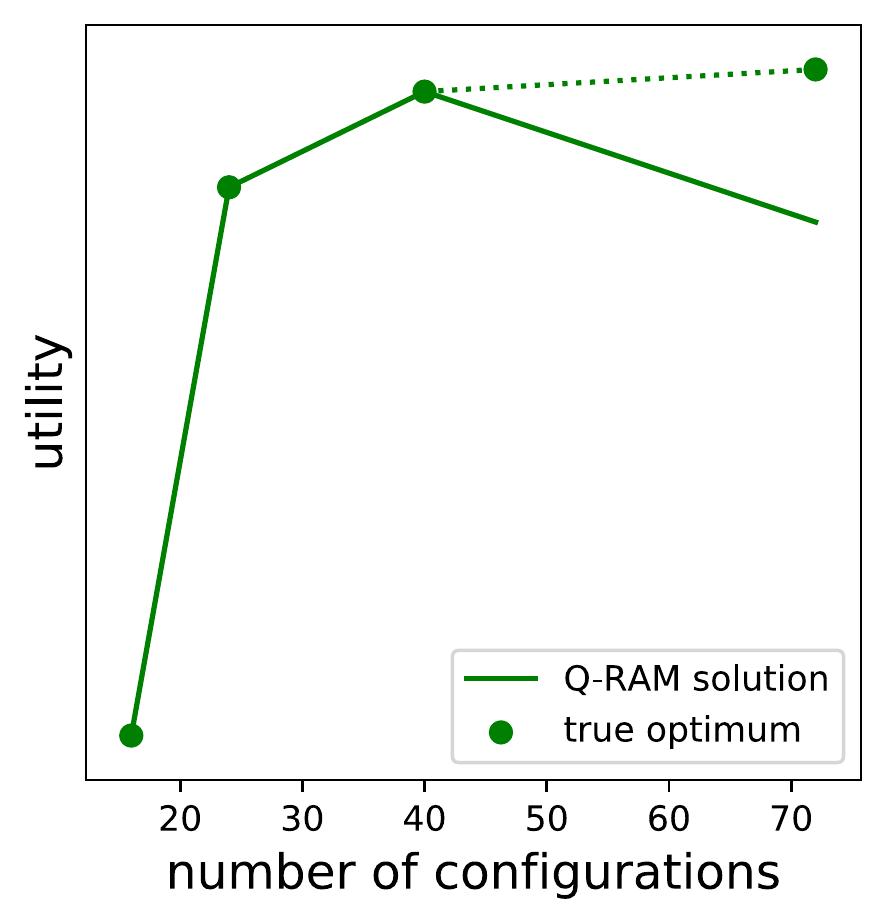}
		\caption{brute force example}
		\label{fig:brute_force}
	\end{subfigure}
	\caption{
		Utility (inversely related to tracking error) of the classical Q-RAM approach by number of configurations per task.\\
	\ref{fig:util_by_configs_A_T}: Varying the discretization of the dwell length and transmit power parameters in the specified scenario.\\
	\ref{fig:brute_force}:  Utility of Q-RAM solution for varying transmit power discretization in a smaller 4 target scenario compared with the true optimum calculated by a brute force search.}
	\label{fig:util_by_configs}
\end{figure}

We apply our solution strategy to the Q-RAM problem as described in \cite{Ghosh2006}.
In particular, the quality, utility and resource functions described there are used.
However, we restrict the problem to tracking tasks (where the quality is inversely related to the tracking error) and allow for 90 possible task configurations in the operational dimensions dwell length, transmit duration and transmit power (see Table~\ref{tab:params}).
A single instance of the scenario consists of a variable number of randomly generated targets of three different types -- helicopters, fighter planes and missiles (see Figure~\ref{fig:scenario_plot} for an example).
\begin{remark}[Dependence of the solution on the configuration space]
	\label{rem:dep_config_space}
Generally, the resulting utility increases with increasing number of possible task configurations and tends to saturate at some point.
However, the classical Q-RAM approach does not always yield higher utilities when refining the discretization of a parameter.
This effect can be observed in Figure~\ref{fig:util_by_configs_A_T} for the varying discretization of the transmit power in the above scenario
and is due to the heuristic nature of the algorithm -- in particular, the restriction to configurations on the boundary of the convex hull in the greedy global optimisation can lead to suboptimal resource usage.
Figure~\ref{fig:brute_force} shows the same phenomenon in a scenario (consisting of 4 targets) small enough to be compared with the actual optimal solution obtained by a brute force search.
Here, the Q-RAM solution with 72 possible configurations per task is worse than the one for a subset of 40 configurations although the true optimum increases.
\end{remark}

\begin{table}[t]
	\centering
	\caption{Parameter selection in the described scenario.}
	\label{tab:params}
	\begin{tabular}[t]{lr}
		\toprule
		&Possible values\\
		\midrule
		dwell length&\numlist{100; 300; 500; 700; 900; 1100} \si{\milli\second}\\
		transmit duration&\numlist{2; 4; 6; 8; 10} \si{\milli\second}\\
		transmit power&\numlist{1;2;4} \si{\kilo\watt}\\
		\bottomrule
	\end{tabular}
\end{table}%

\subsection{Agent architecture}

\begin{figure*}[htb] 
	\centering
	\includegraphics[width=.98\textwidth]{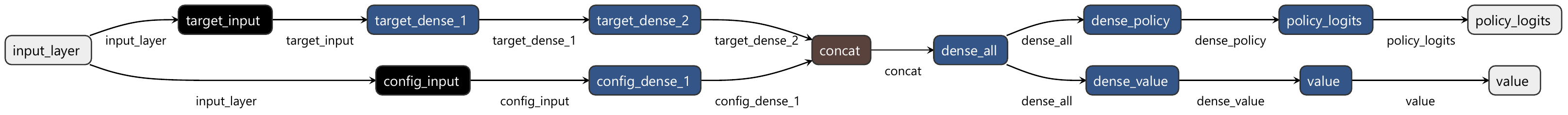}
	\caption{The actor critic network, visualisation by Netron (\url{https://github.com/lutzroeder/netron}).}
	\label{fig:network_architectures}
\end{figure*}

The agent uses a single-worker advantage actor critic network (A2C, a synchronous version of asynchronous advantage actor critic networks \cite{A3C}, see \cite{RingBlog} for an implementation).
To showcase the functionality of the general idea described in Section~\ref{sec:our_solution}, the network architecture itself is
a straight-forward A2C-architecture with 100 neurons per layer (see Figure~\ref{fig:network_architectures}).
The only speciality is the split of the input data into situational, i.e.\ target-related data and the given task configuration. These components are concatenated after running through two (one, respectively) dense layers. All layers except for the output layers use rectified linear units as activation functions.

\subsection{Training}

The agent is trained by reinforcement learning via interactions with random instances of the environment as described in Section~\ref{sec:our_solution}. In reinforcement learning it is common to use a factor of about 0.99 to discount future rewards.
In contrast, we chose a very low discount factor of 0.005 and limited the number of interactions with the environment to three steps per episode.
The reason lies in the design of the particular environment. If future rewards are not sufficiently discounted, the agent can generate value by repeating actions along suitable triangles in resource-utility-space.
Training was performed using RMSprop \cite{HintonCoursera} and with capping rewards for better stability.

\subsection{Performance}

\begin{figure}[htb] 
	\centering
	\includegraphics[width=.4\textwidth]{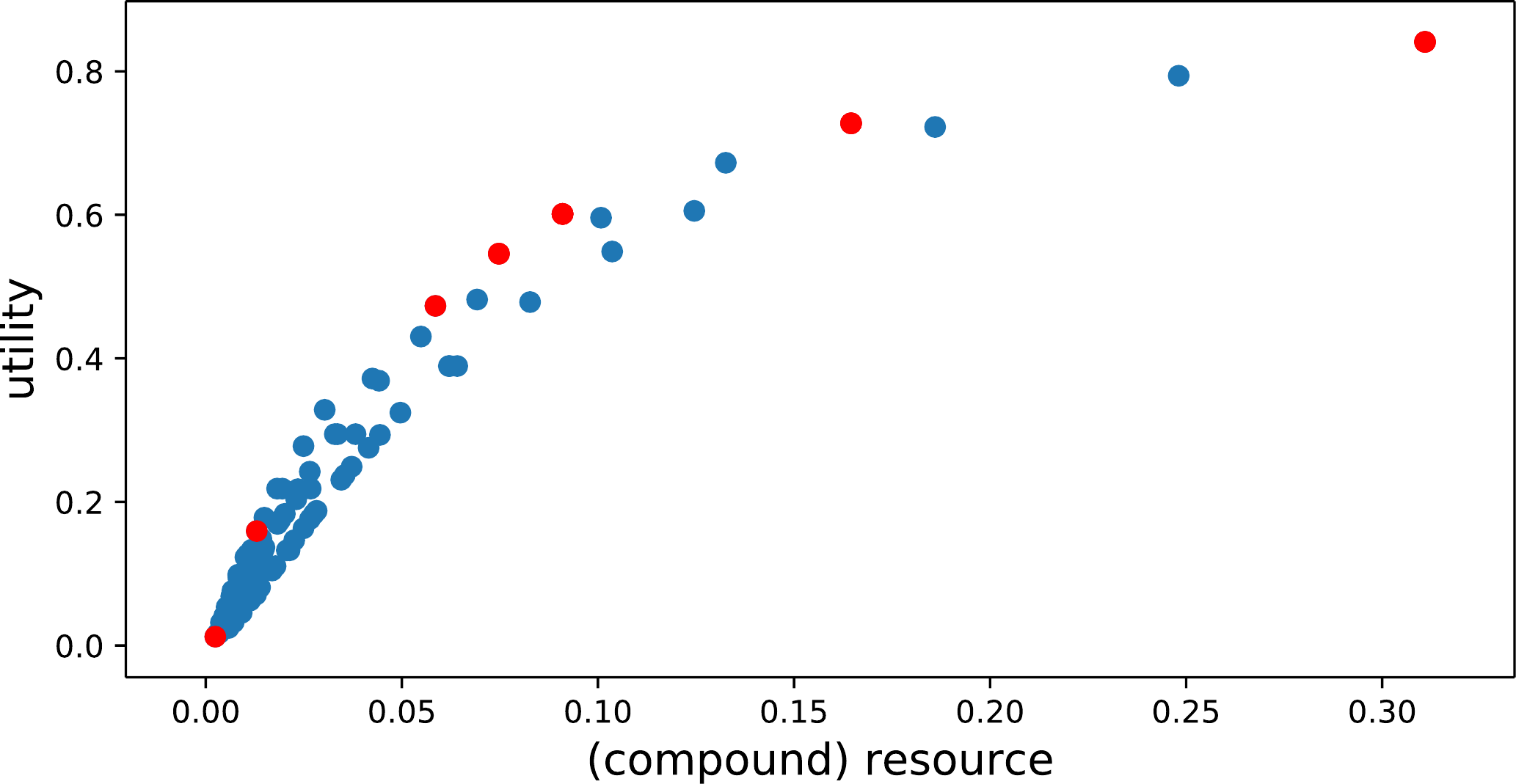}
	\caption{Possible configurations of a single tracking task embedded into resource-utility-space. Configurations chosen by the network agent are marked in red.}
	\label{fig:setpoints}
\end{figure}

The agent is successful in learning to choose \emph{good} task configurations, i.e.\ chosen configurations yield a high utility for their respective resource requirements. Figure~\ref{fig:setpoints} shows all possible task configurations for a single tracking task. The ones selected by the neural network are marked in red and are \emph{good} in the above sense, as they are on or close to the upper boundary of the convex hull of all points.

\begin{figure}[htb] 
	\centering
	\includegraphics[width=.4\textwidth]{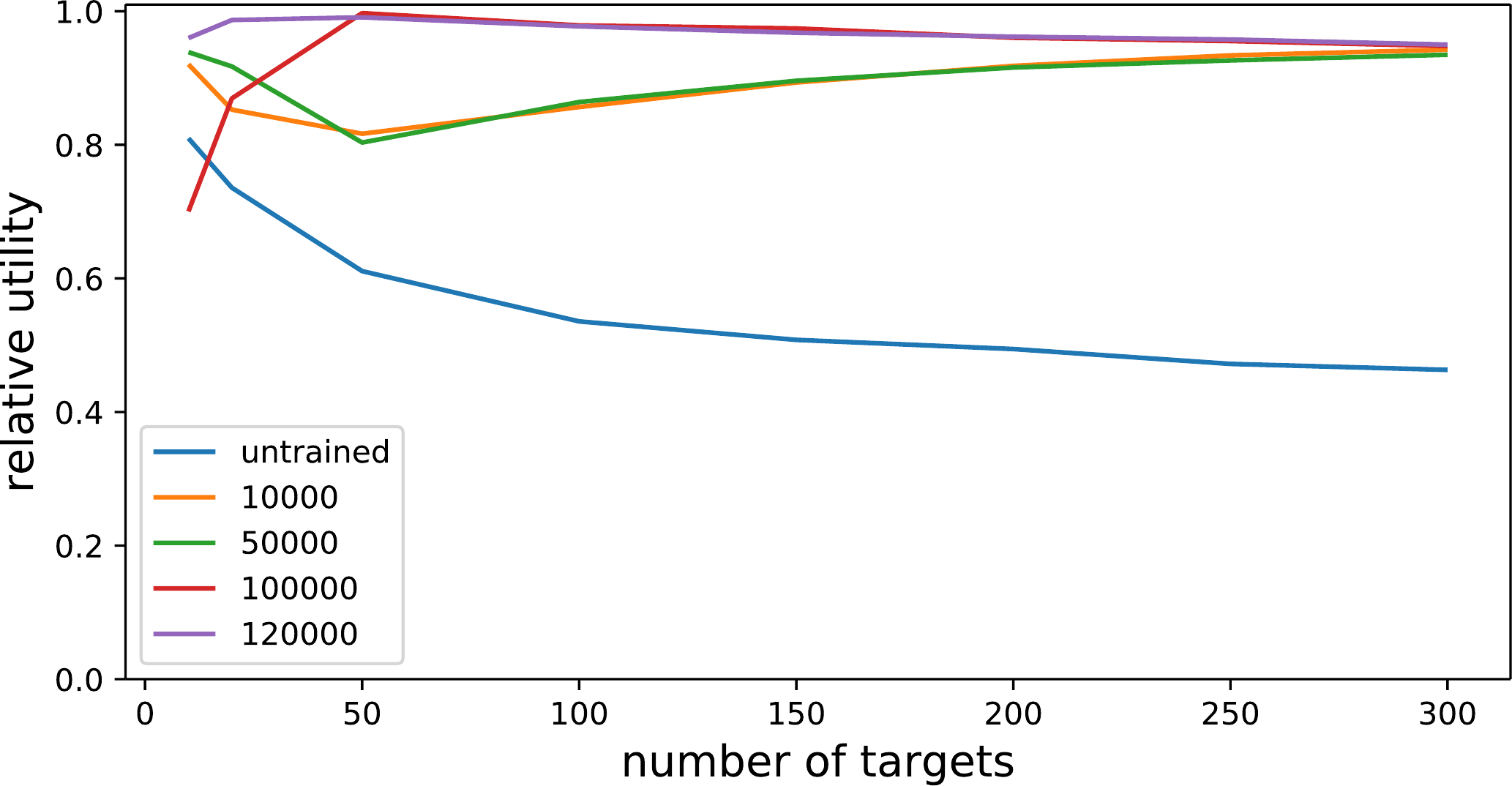}
	\caption{Comparison of neural network agents with full Q-RAM solution for various training steps (utility averaged over 20 random runs) in the described scenario.}
	\label{fig:comparison}
\end{figure}

To apply the trained neural network to the global Q-RAM problem, we use the implementation described in Section~\ref{sec:our_solution}. The performance of the resulting resource management module as compared with the full classical solution is displayed in Figure~\ref{fig:comparison}.
An agent that was trained for 120,000 steps reaches a utility between 97\% to 99\% when compared to the classic Q-RAM solution in the relevant range of 20 to 150 simultaneous targets to track.
Utility increases with training time and decreases slightly with increasing number of targets to be tracked.
In addition to the possibility that random target data might be far off from examples experienced during training, 
an explanation for this decline might be that the selection of good configurations per task is not fine enough.
However, the results suggest that this problem can be mitigated by a longer training period.

\begin{figure}[htb] 
	\centering
	\includegraphics[width=.4\textwidth]{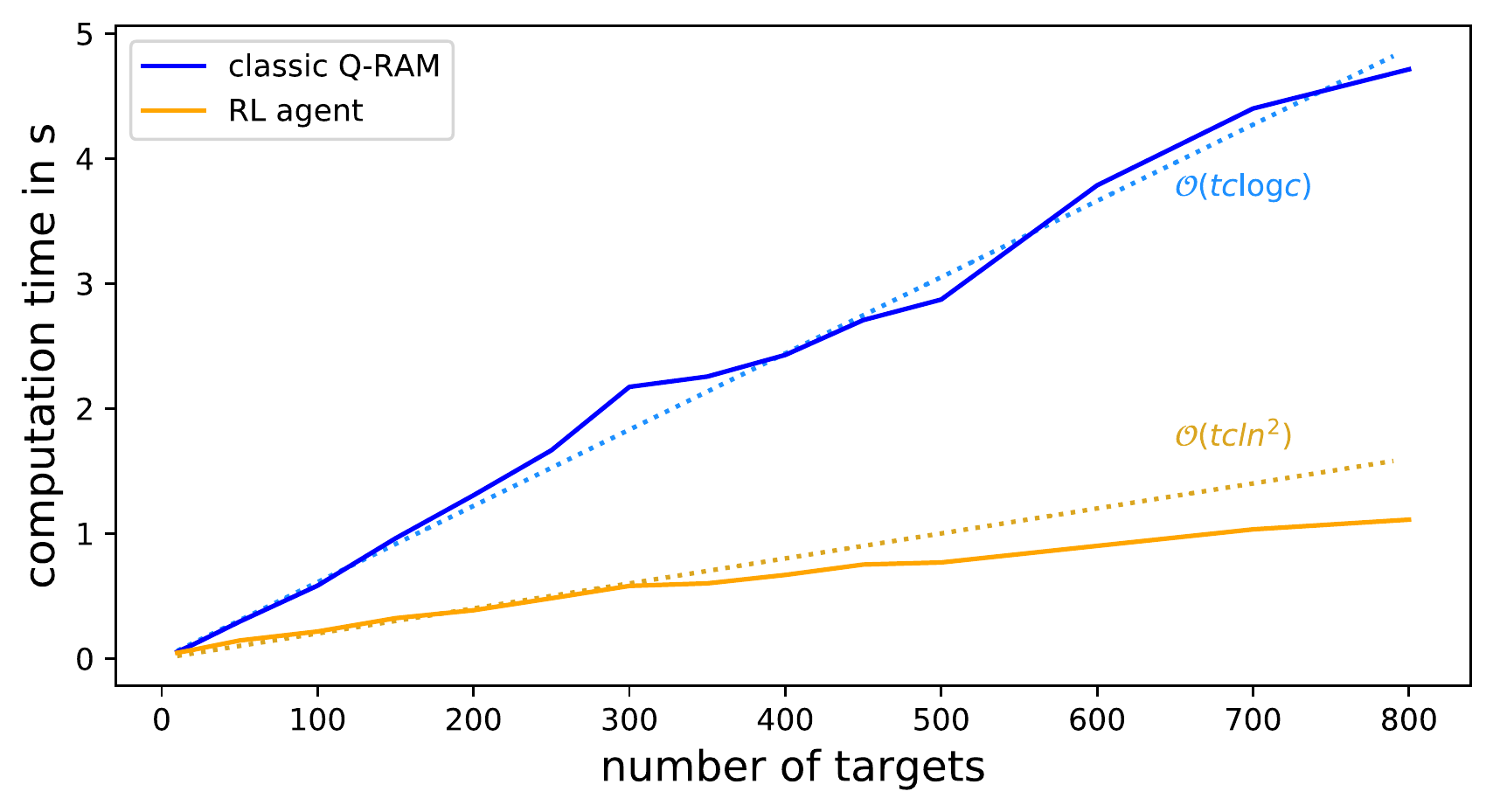}
	\caption{Computation times (experimental values) of trained network agent and full Q-RAM solution for a fixed number of $c=90$ possible configurations per task in the described scenario (averaged over 20 random runs).}
	\label{fig:runtime}
\end{figure}

\begin{figure}[htb] 
	\centering
	\includegraphics[width=.4\textwidth]{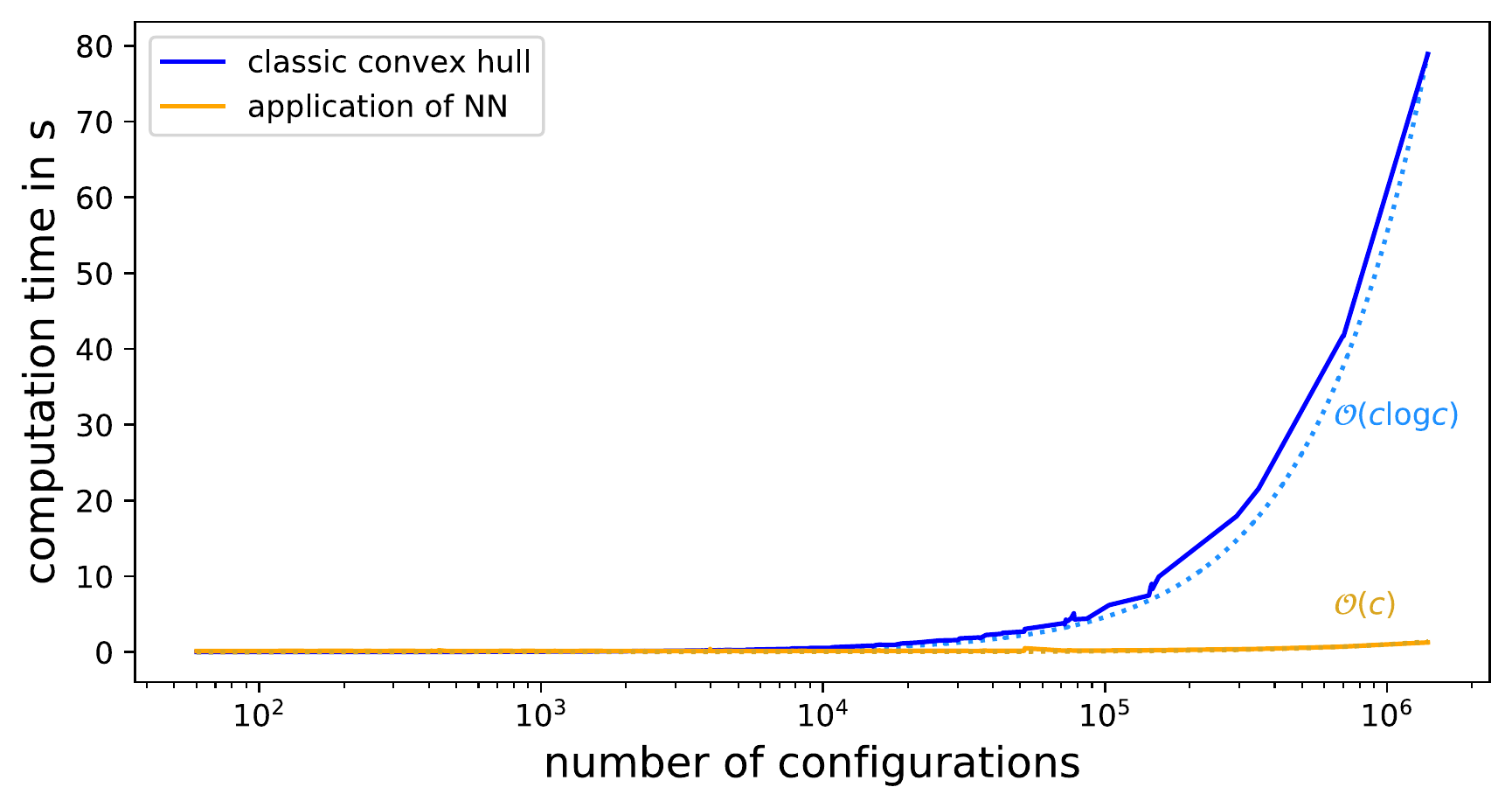}
	\caption{Computation of the Q-RAM job list of a single task with the classic convex hull approach and single application of the neural network by number of possible task configurations (average over 20 random runs).}
	\label{fig:runtime_qram}
\end{figure}

The advantages of the neural network based RL agent become apparent in Figures~\ref{fig:runtime} and~\ref{fig:runtime_qram}. Even for the relatively small problem under consideration with a total of only 90 possible configurations per task, the neural network considerably outperforms the classical approach in computation time.
Even more so in a more general setting, as the computation time of the convex hull operation in the classic Q-RAM approach increases with increasing number of possible configurations per task, whereas the duration of a single feed-forward pass of the net remains almost constant.

\begin{remark}
The actual computations were performed on a Lenovo Thinkpad L590 with an Intel i7-8565U processor using Python and (non-GPU) Tensorflow.
\end{remark}

\begin{figure}[tb] 
	\centering
	\includegraphics[width=.4\textwidth]{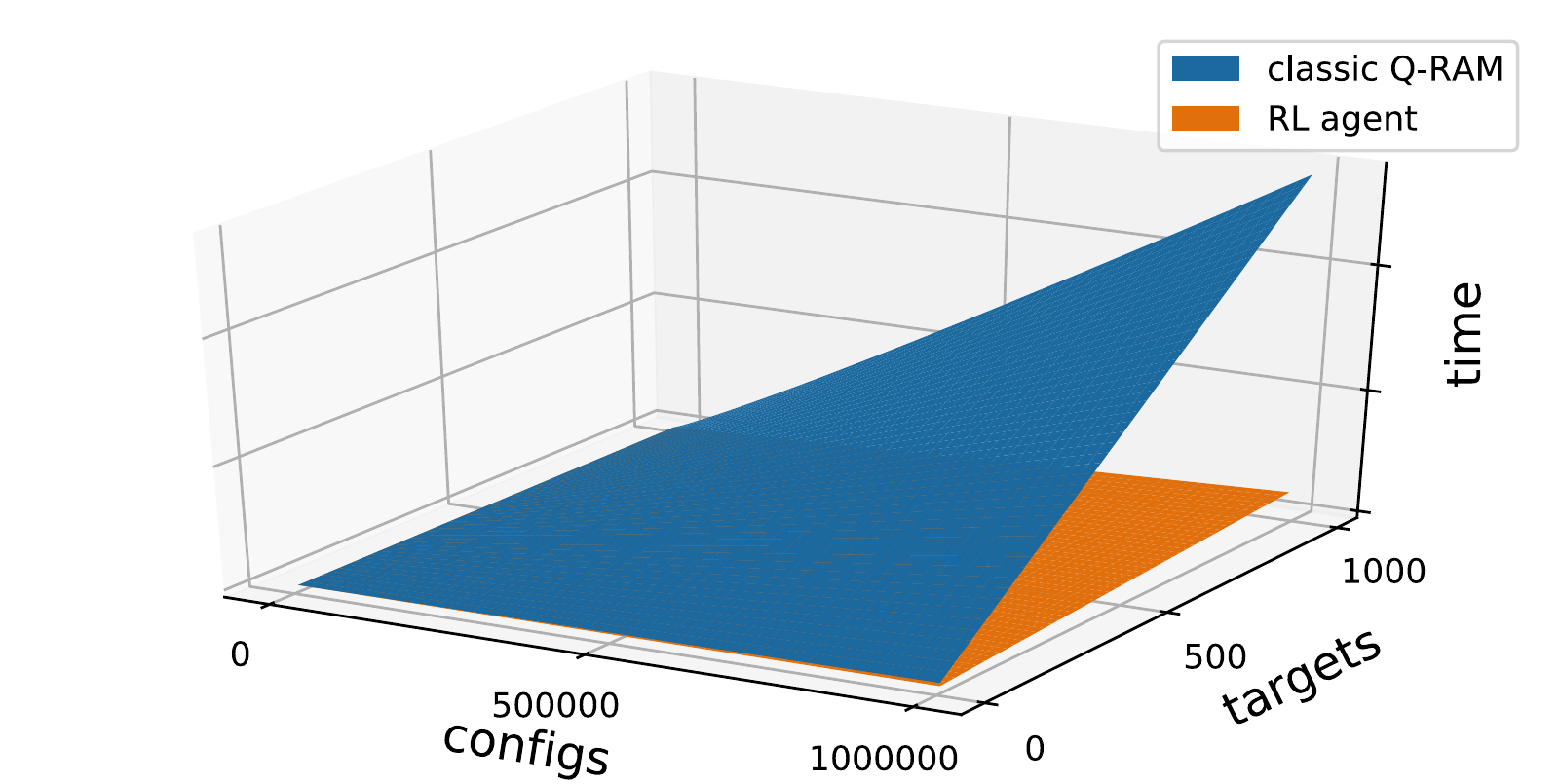}
	\caption{Theoretic computation time by computational complexity of the classical Q-RAM solution and the RL agent based resource management.}
	\label{fig:complexity_3d}
\end{figure}

These experimental results are in line with the asymptotic behaviours of the algorithms. Both the classical Q-RAM solution and the trained RL agent based resource management module perform worst in a scenario where all task configurations lie in the boundary of their convex hull in resource-utility-space and the global optimiser does not stop resource allocation early because of an abundance of resource.
In this case, the classical solution has a computational complexity of $\mathcal{O}(tc\log c)$, with $c$ the number of possible configurations per task and $t$ the number of targets.
The neural network agent including the global optimisation as described in Section~\ref{sec:method2} possesses a complexity of $\mathcal{O}(tcln^2)$, where $l$ is the number of neural network layers and $n$ the number of neurons per layer. As $l$ and $n$ are both fixed after choosing an appropriate architecture, the resulting agent has a computational complexity of $\mathcal{O}(tc)$ and thus improves the classical approach by a factor of $\log c$ (see Figure~\ref{fig:complexity_3d}). Notice that in this computationally worst case, the resulting system utilities are equal for both solution strategies.

\begin{figure}[htb] 
	\centering
	\includegraphics[width=.4\textwidth]{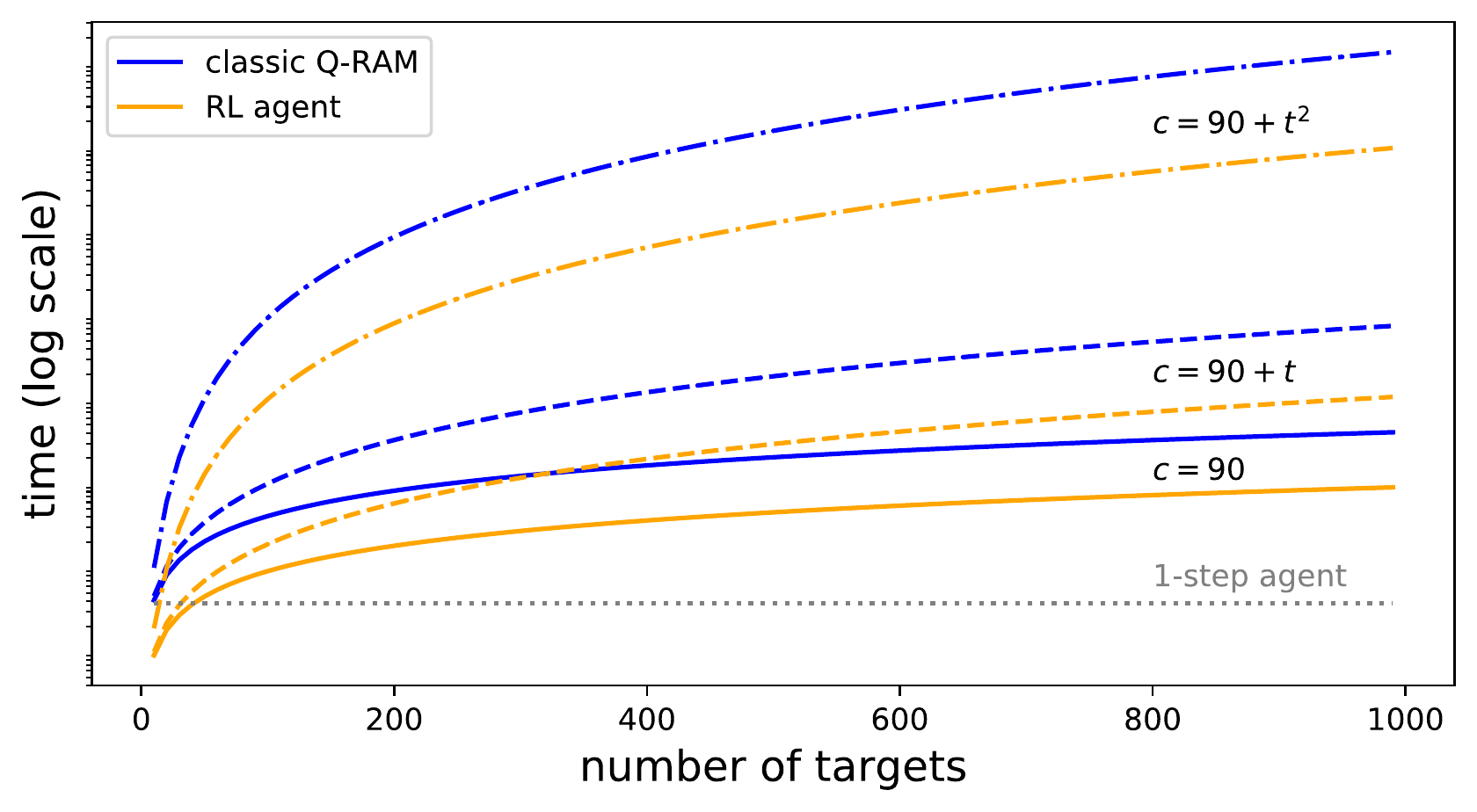}
	\caption{Theoretic computation time by computational complexity of the classical Q-RAM solution and the RL agent by number of targets $t$ with increasing number of possible configurations $c$ per task. The dotted grey line represents a neural network agent performing the global resource allocation in a single step.}
	\label{fig:complexity_2d}
\end{figure}

It seems plausible that in a real world cognitive radar, the number of possible configurations per task is not necessarily fixed but rather increases with increasing target number in field of view, as this allows for a finer resource allocation and also completely new configurations e.g.\ realising the tracking of multiple targets with the same beam.
The corresponding impact on performance is depicted in Figure~\ref{fig:complexity_2d},
where the growth in the number of possible task configurations $c$ is assumed linear or quadratic in the number of targets $t$ and the results are shown next to the constant $c$ reference of $90$ configurations.
The computation time is based on the computational complexity in the worst case scenario as described above.

\subsection{Scalability}

As shown above, the RL agent based radar resource management
scales nicely with number of tasks, i.e.\ in contrast with the classical Q-RAM approach, our implementation does not lead to an extreme computational complexity in overload situations.
Clearly, if the number of configurations per task becomes too high, and in real world applications it can easily go into the millions, a standard network architecture with one output neuron per configuration is not feasible.
Therefore we propose to design the agent using 
a Wolpertinger architecture \cite{Dulac2015}, which has been successfully applied to problems with large action spaces to replace actor critic or deep Q-networks.

A natural next step is to train an agent to directly solve the global optimisation problem and, in a further advancement, to directly output a scheduled radar timeline. These approaches are currently under investigation by the authors. Nevertheless, alluring as these holistic solutions might be, the presented method has some practical benefits -- namely the already mentioned possibility of training on a per task type basis,
enabling the operator to value quality indicators and thus setting goals per mission,
and to quickly enhance the system's capabilities by combining the trained agent with classical solutions for tasks not considered during training.

\section{Conclusion}
\label{sec:conclusion}

An environment has been designed to enable a neural network agent to learn desirable task configurations in the radar Q-RAM setting via deep reinforcement learning. The agent is incorporated into a global optimisation scheme to solve the Q-RAM problem. As a proof of concept, the algorithm was applied to a radar tracking problem and the results were compared with the classical Q-RAM strategy.

The proposed solution drastically improves computation time by avoiding the need for task configurations to be embedded into resource-utility-space without introducing the disadvantages of existing approaches to mitigate performance problems as described in Section~\ref{sec:classical_solution}.

Being based on deep reinforcement learning, the solution has the potential to make use of more complex and realistic performance models since no domain knowledge (e.g.\ a priori knowledge about the environment or expected rewards) is required in the agent.
Furthermore, training the network by reinforcement learning does not involve solving the Q-RAM problem by existing, cost-intensive methods to generate labelled training data.
The RL agent based approach can be easily integrated into existent Q-RAM implementations and enables a system to evolve and learn post-mission, paving the way for a truly cognitive radar.

\section*{Acknowledgements}
The authors are grateful to Christoph Fischer and Hensoldt Sensors GmbH for fruitful discussions.


\bibliographystyle{IEEEtran}
\bibliography{IEEEabrv,lit}

\begin{thebibliography}{10}
\providecommand{\url}[1]{#1}
\csname url@samestyle\endcsname
\providecommand{\newblock}{\relax}
\providecommand{\bibinfo}[2]{#2}
\providecommand{\BIBentrySTDinterwordspacing}{\spaceskip=0pt\relax}
\providecommand{\BIBentryALTinterwordstretchfactor}{4}
\providecommand{\BIBentryALTinterwordspacing}{\spaceskip=\fontdimen2\font plus
\BIBentryALTinterwordstretchfactor\fontdimen3\font minus
  \fontdimen4\font\relax}
\providecommand{\BIBforeignlanguage}[2]{{%
\expandafter\ifx\csname l@#1\endcsname\relax
\typeout{** WARNING: IEEEtran.bst: No hyphenation pattern has been}%
\typeout{** loaded for the language `#1'. Using the pattern for}%
\typeout{** the default language instead.}%
\else
\language=\csname l@#1\endcsname
\fi
#2}}
\providecommand{\BIBdecl}{\relax}
\BIBdecl

\bibitem{Haykin2006}
S.~Haykin, ``Cognitive radar: a way of the future,'' \emph{IEEE Signal
  Processing Magazine}, vol.~23, pp. 30--40, 2006.

\bibitem{Raj1997}
R.~Rajkumar, C.~Lee, J.~Lehoczky, and D.~Siewiorek, ``A resource allocation
  model for {Q}o{S} management,'' in \emph{Proceedings of the IEEE Real-Time
  Systems Symposium}, 1997, pp. 298--307.

\bibitem{Ghosh2004}
S.~Ghosh, J.~Hansen, R.~Rajkumar, and J.~Lehoczky, ``Adaptive {Q}o{S}
  optimizations with applications to radar tracking,'' in \emph{10th
  International Conference on Real-Time and Embedded Computing Systems and
  Applications (RTCSA)}, 2004.

\bibitem{Ghosh2006}
S.~Ghosh, R.~Rajkumar, J.~Hansen, and J.~Lehoczky, ``Integrated {Q}o{S}-aware
  resource management and scheduling with multi-resource constraints,''
  \emph{Real-Time Syst.}, vol.~33, pp. 7--46, 2006.

\bibitem{Lee1998}
C.~Lee and D.~Siewiorek, ``An approach for quality of service management,''
  Technical Report CMU-CS-98-165, Computer Science Department, Carnegie Mellon
  University, Tech. Rep., 1998.

\bibitem{Lee1999}
C.~Lee, J.~Lehoczky, R.~Rajkumar, and D.~Siewiorek, ``On quality of service
  optimization with discrete {Q}o{S} options,'' in \emph{Proceedings of the
  IEEE Real-time Technology and Applications Symposium}, 1999, pp. 276--286.

\bibitem{CharlishPhD}
A.~Charlish, ``Autonomous agents for multi-function radar resource
  management,'' Ph.D. dissertation, University College London, London, 2011.

\bibitem{MooDing}
P.~Moo and Z.~Ding, \emph{Adaptive Radar Resource Management}.\hskip 1em plus
  0.5em minus 0.4em\relax London: Academic Press, 2015.

\bibitem{Mao2016}
H.~Mao, M.~Alizadeh, I.~Menache, and S.~Kandula, ``Resource management with
  deep reinforcement learning,'' in \emph{HotNets '16: Proceedings of the 15th
  ACM Workshop on Hot Topics in Networks}, 2016, pp. 50--56.

\bibitem{Shaghaghi2018}
M.~Shaghaghi, R.~S. Adve, and Z.~Ding, ``Multifunction cognitive radar task
  scheduling using monte carlo tree search and policy networks,'' \emph{IET
  Radar, Sonar \& Navigation}, vol.~12, pp. 1437--1447, 2018.

\bibitem{Dong2020}
R.~Dong, C.~She, W.~Hardjawana \emph{et~al.}, ``Deep learning for radio
  resource allocation with diverse quality-of-service requirements in 5{G},''
  arXiv:2004.00507.

\bibitem{Ak2019}
S.~Ak and S.~Br\"uggenwirth, ``Avoiding jammers: A reinforcement learning
  approach,'' arXiv:1911.08874.

\bibitem{Kang2018}
L.~Kang, J.~Bo, L.~Hongwei, and L.~Siyuan, ``Reinforcement learning based
  anti-jamming frequency hopping strategies design for cognitive radar,'' in
  \emph{2018 IEEE International Conference on Signal Processing, Communications
  and Computing (ICSPCC)}, Qingdao, 2018, pp. 1--5.

\bibitem{Liu2019}
P.~Liu, Y.~Liu, T.~Huang \emph{et~al.}, ``Cognitive radar using reinforcement
  learning in automotive applications,'' arXiv:1904.10739.

\bibitem{Ahmed2020}
A.~M. Ahmed, A.~A. Ahmad, S.~Fortunati \emph{et~al.}, ``Reinforcement learning
  based beamforming for massive mimo radar multi-target detection,''
  arXiv:2005.04708.

\bibitem{SuttonBarto}
R.~Sutton and G.~Barto, \emph{Reinforcement Learning: An Introduction}.\hskip
  1em plus 0.5em minus 0.4em\relax Cambridge, MA: MIT Press, 2018.

\bibitem{Mnih2015}
V.~Mnih, K.~Kavukcuoglu, D.~Silver \emph{et~al.}, ``Human-level control through
  deep reinforcement learning,'' \emph{Nature}, vol. 518, pp. 529--533, 2015.

\bibitem{Lillicrap2015}
T.~Lillicrap, J.~Hunt, A.~Pritzel \emph{et~al.}, ``Continuous control with deep
  reinforcement learning,'' arXiv:1509.02971.

\bibitem{Dulac2015}
G.~Dulac-Arnold, R.~Evans, H.~van Hasselt \emph{et~al.}, ``Deep reinforcement
  learning in large discrete action spaces,'' arxiv:1512.07679.

\bibitem{A3C}
V.~Mnih, A.~{Puigdomènech Badia}, M.~Mirza \emph{et~al.}, ``Asynchronous
  methods for deep reinforcement learning,'' arXiv:1602.01783.

\bibitem{RingBlog}
\BIBentryALTinterwordspacing
R.~Ring, ``Deep reinforcement learning with tensorflow 2.1,'' Blog post.
  [Online]. Available:
  \url{http://inoryy.com/post/tensorflow2-deep-reinforcement-learning/}
\BIBentrySTDinterwordspacing

\bibitem{HintonCoursera}
\BIBentryALTinterwordspacing
G.~Hinton, ``Neural networks for machine learning,'' Lecture course, University
  of Toronto / Coursera, 2012. [Online]. Available:
  \url{https://www.cs.toronto.edu/~hinton/coursera_slides.html}
\BIBentrySTDinterwordspacing

\end{thebibliography}

\end{document}